\documentclass[preprint, 12pt]{elsarticle}
\newcommand{\ket}[1]{|{#1}\rangle}
\newcommand{\bra}[1]{\langle{#1}|}

\newcommand{\be}{\begin{equation}}
\newcommand{\ee}{\end{equation}}

\usepackage{amsfonts}
\usepackage{amsmath}
\usepackage{rotating}
\usepackage{graphicx}
\usepackage{graphics}
\usepackage{epsfig}
\usepackage{amsthm}
\usepackage{amssymb}
\usepackage{amsmath}
\usepackage{amssymb}
\usepackage{graphicx}
\usepackage{lscape}
\usepackage{color}
\usepackage{epstopdf}

\journal{Somewhere}

\begin{document}

\begin{frontmatter}

\title{Fermionic Luttinger liquids from a microscopic perspective}

\author{Manuel Valiente}
\address{SUPA, Institute of Photonics and Quantum Sciences, Heriot-Watt University, Edinburgh EH14 4AS, United Kingdom}
\author{Lawrence G. Phillips}
\address{SUPA, Institute of Photonics and Quantum Sciences, Heriot-Watt University, Edinburgh EH14 4AS, United Kingdom}
\author{Nikolaj T. Zinner}
\address{Department of Physics and Astronomy, Aarhus University, Ny Munkegade 120, 8000 Aarhus C, Denmark}
\author{Patrik {\"O}hberg}
\address{SUPA, Institute of Photonics and Quantum Sciences, Heriot-Watt University, Edinburgh EH14 4AS, United Kingdom}
\begin{abstract}
We consider interacting one-dimensional, spinless Fermi gases, whose low-energy properties are described by Luttinger liquid theory. We perform a systematic, in-depth analysis of the relation between the macroscopic, phenomenological parameters of Luttinger liquid effective field theory, and the microscopic interactions of the Fermi gas. In particular, we begin by explaining how to model effective interactions in one dimension, which we then apply to the main forward scattering channel -- the interbranch collisions -- common to these systems. We renormalise the corresponding interbranch phenomenological constants in favour of scattering phase shifts. Interestingly, our renormalisation procedure shows (i) how Luttinger's model arises in a completely natural way -- and not as a convenient approximation -- from Tomonaga's model, and (ii) the reasons behind the interbranch coupling constant remaining unrenormalised in Luttinger's model. We then consider the so-called intrabranch processes, whose phenomenological coupling constant is known to be fixed by charge conservation, but whose microscopic origin is not well understood. We show that, contrary to general belief and common sense, the intrabranch interactions appearing in Luttinger liquid theory do not correspond to an intrabranch scattering channel, nor an energy shift due to intrabranch interactions, in the microscopic theory. Instead, they are due to interbranch processes. We finally apply our results to a particular example of an exactly solvable model, namely the fermionic dual to the Lieb-Liniger model in the Tonks-Girardeau and super-Tonks-Girardeau regimes. 
\end{abstract}

\begin{keyword}
Luttinger liquids \sep Renormalisation \sep Effective field theory \sep One-dimensional systems
\end{keyword}

\end{frontmatter}

%%%%%%%%%%%%%Introduction%%%%%%%%%%%%%%
\section{Introduction}
Interacting many-body systems in one spatial dimension, once considered theoretical playgrounds for the more involved ``physical'' three-dimensional systems, hold currently, and for quite some time, the status of physically relevant theories. Effective reduced-dimensional systems are indeed routinely prepared and studied experimentally. Prominent examples include ultracold atomic gases \cite{ref1,ref2,ref3,ref4,ref5,ref6,ref7}, organic conductors \cite{ref8,ref9,ref10} and nanotubes \cite{ExperimentNanotubes}. From the theoretical point of view, one-dimensional systems are very appealing, for a number of reasons. Firstly, there are a number of exactly solvable models in one spatial dimension. Examples include those models solved via the Bethe ansatz technique \cite{Bethe}, such as the spinless Bose gas with Dirac delta interactions (Lieb-Liniger model) \cite{LiebLiniger} and its spin-1/2 fermionic counterpart \cite{Yang}, and models with unbroken SuSy, which admit Jastrow-product ground state wave functions, such as Sutherland's model with inverse-square interactions and variations thereof \cite{Sutherland}. More recently, strongly-coupled mesoscopic multicomponent systems in arbitrary external potentials have been the subject of intense investigation \cite{Multicomponent1,Multicomponent2,Multicomponent3,Multicomponent4}. Moreover, some numerical methods work especially well for one-dimensional systems. For instance, the density matrix renormalization group (DMRG) \cite{DMRG} is extremely powerful and flexible to use in most one-dimensional many-body problems on a lattice. 

The low-energy properties of many of the above systems can be described by the universality class of Luttinger liquids \cite{Giamarchi,Tsvelik}. In brief, the properties of these systems, such as the collective excitation spectrum, correlation functions, or density of states, at energies close to the Fermi energy, exhibit universal behaviour that is described by the exact solution of Tomonaga-Luttinger's model \cite{LuttingerModel1,LuttingerModel2} via bosonization \cite{LiebMattis,LutherPeschel,Mattis,Coleman,Mandelstam}. For instance, the density of states behaves as a power law, where the exponent and the proportionality constant are the only microscopic, non-universal details, around the Fermi points \cite{DeNijs,Dzy,Efetov}.

Historically, there are two fundamentally different, yet complementary approaches for the introduction of Luttinger Liquid phenomenological models for one-dimensional quantum gases. The first one, called constructive bosonisation \cite{Delft}, naturally arises for fermionic systems upon linearisation of the kinetic energy dispersion, as introduced by Tomonaga \cite{LuttingerModel1}. The second, more phenomenological, and also more general approach, is field-theoretical bosonisation, introduced by Haldane \cite{HaldaneJPC}. 

In this article, we consider one-dimensional spinless Fermi gases in the fermionic Luttinger liquid regime, that is, when their low-energy behaviour can be modelled by Luttinger's model. We will follow the constructive, rather than the field-theoretical route, which is most convenient for weakly-interacting fermions. In particular, we study the microscopic origin of the interaction coupling constants in Luttinger's model. We first address the main forward-scattering channel in the microscopic theory, corresponding to the collision of two fermions with incident momenta close to the two opposite Fermi points. In the very weak-coupling regime, these interbranch processes are well understood, and we show how to renormalise Luttinger's interbranch interaction in favour of scattering data (i.e. the two-body phase shift). From our analysis of the interbranch collision channel, we show how Luttinger's model naturally arises from Tomonaga's model upon renormalisation of the Fermi points, which are to be considered as bare parameters of the theory. In this way, we also explain the reason behind the interbranch coupling constant remaining "unrenormalised" in Luttinger's model, a well known fact, as opposed to Tomonaga's model, where the theory requires non-trivial renormalisation. We then consider intrabranch interactions, i.e. interactions between particles moving in the same direction. In this case, symmetry dictates that the intrabranch coupling constant must be identical to the interbranch one \cite{BruusFlensberg}. However, the current understanding of the microscopic origin of the intrabranch interaction in Luttinger's model is quite unsatisfactory. Usually \cite{BruusFlensberg,Eggert}, its origin is said to be linked to the forward scattering of fermions near the same Fermi point, i.e. in the same branch. Even though this picture sounds reasonable, and easy to grasp physically, it has some major issues. Here, we carefully illustrate these issues, and find that the effective {\it intra}branch interaction in Luttinger's model corresponds, in the microscopic fermionic model, to an energy shift due to {\it inter}branch interactions. We finally apply our results in order to obtain the Luttinger's parameter and the speed of excitations in the Tonks-Girardeau regime \cite{Girardeau} of the Lieb-Liniger gas, corresponding to the weak-coupling limit of its fermionic dual \cite{CheonShigehara}.  

%%%%%%%%%%%%Microscopic system%%%%%%%%%%%%%%%%%%%%%%%
\section{Microscopic system}
We consider a system of $N$ non-relativistic spinless fermions of mass $m$ interacting via pairwise potentials $W$ in one spatial dimension (1D). The Hamiltonian in the first quantisation has the form
\begin{equation}
H=-\frac{\hbar^2}{2m}\sum_{i=1}^N \frac{\partial^2}{\partial x_i^2}+\sum_{i<j=1}^N W(x_i-x_j),
\end{equation}
where $x_i$ is the position of the $i$-th particle. In the momentum representation, the two-body interaction is given by
\begin{equation}
V(k,k')\equiv V(q) = \int_{-\infty}^{\infty}dx e^{iqx} W(x),
\end{equation}
where we have defined $q=k-k'$. With this notation, we can write down the second-quantised Hamiltonian in the momentum representation,
\begin{equation}
H=\sum_{k}\epsilon(k) c_k^{\dagger}c_k + \frac{1}{2L}\sum_{kk'q}V(q)c_{k+q}^{\dagger}c_{k'-q}^{\dagger}c_{k'}c_k.\label{Hamiltonianmomentum}
\end{equation}
Above, $\epsilon(k)=\hbar^2k^2/2m$ is the single-particle kinetic energy dispersion, $L$ is the system's size, $c_k$ ($c_k^{\dagger}$) annihilates (creates) a fermion with momentum $k$, and therefore satisfies canonical anticommutation relations
\begin{equation}
\{c_k,c_{k'}^{\dagger}\}=\delta_{k,k'}.
\end{equation}
In Hamiltonian (\ref{Hamiltonianmomentum}) we have assumed periodic boundary conditions, that is, $k=2\pi n/L$ with $n$ being integer numbers.

Since we are working with fermions, it is convenient to introduce even and odd wave components in the interaction as $V=V_s+V_p$. The even (odd) wave component, acting only on bosonic (fermionic) wave functions, sometimes called "s-wave" ("p-wave"), is given by the following symmetric (antisymmetric) combination 
\begin{align}
V_s(k,k')&=\frac{1}{2}\left[V(k,k')+V(k,-k')\right],\\
V_p(k,k')&=\frac{1}{2}\left[V(k,k')-V(k,-k')\right].\label{pwave}
\end{align}
In the following, we will use $V$ or $V_p$ at our discretion, depending on whichever is more convenient in each situation.

%%%%%%%%%%%Low-energy interactions in vacuum%%%%%%%%%%%%%%%%%%%%%%%%
\section{Model interactions}
In many applications, it is advantageous to construct effective interactions that mimic the exact scattering properties of a system at the relevant energy scales. This can greatly simplify the calculations while keeping a correct description of collisional processes, and allow for universal (i.e. independent of microscopic details) predictions of many quantities of interest. For instance, low-energy scattering is adequately described by means of "pionless" effective field theory (EFT) \cite{WeinbergEFT}, in which the interaction is Taylor-expanded around $q=k-k'=0$ as
\begin{equation}
V(k,k')=g_a + g_b (k-k')^2+\ldots .
\end{equation}
The lowest-order contribution to the p-wave potential above is given by
\begin{equation}
V_p(k,k')\approx-2g_b k k' \equiv \tilde{g}kk'\equiv V_p^{(4)}(k,k'),\label{V4}
\end{equation}
where $\tilde{g}$ is the bare coupling constant that is to be renormalised in favour of low-energy scattering data. 

\begin{figure}[t]
\includegraphics[width=1\textwidth]{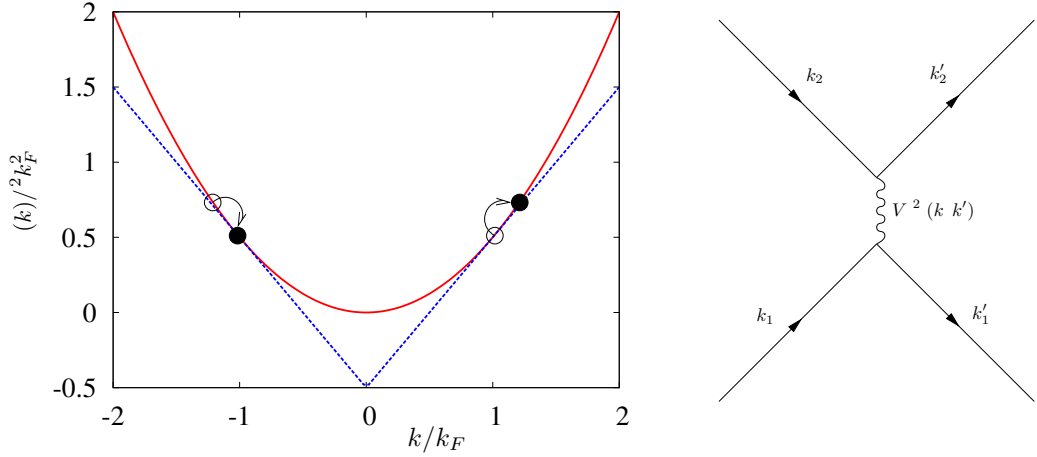}
\caption{Left: depiction of interbranch scattering processes around the Fermi points. The red solid line corresponds to the Galilean (parabolic) dispersion, and the blue dashed line corresponds to the linearised approximation around the Fermi points. Right: interbranch interaction diagram for spinless fermions.}
\label{fig:g2-process}
\end{figure}

In this article, we shall be interested in collisions of particles with relative momenta of $\mathcal{O}(k_F)$, where $k_F$ is the Fermi momentum. The EFT interaction (\ref{V4}) can be used if the particles have small relative momenta $kk'/k_F^2\ll 1$. If, instead, $kk'=\mathcal{O}(k_F^2)$, the lowest order interaction is given by
\begin{equation}
V_p(\pm k_F,\pm k_F)\approx \frac{1}{2}\left[V(0)-V(2k_F)\right].\label{Vp1}
\end{equation}
The resulting low-energy interaction is sometimes referred to as "constant" \cite{Giamarchi}. This is a misconception that can lead to contradictions \cite{BruusFlensberg}. If we wish to use Eq. (\ref{Vp1}) for all momenta, this must satisfy $V_p(k,-k')=-V_p(k,k')$ (see Eq. (\ref{pwave})). This gives a model interaction of the form
\begin{align}
V_p(k,k')&\approx\frac{1}{2}\left[V(0)-V(2k_F)\right]\mathrm{sgn}(k)\mathrm{sgn}(k')\nonumber \\
&\equiv \tilde{g}_2\mathrm{sgn}(k)\mathrm{sgn}(k')\equiv V_p^{(2)}(k,k').\label{Vp2}
\end{align}
The collisions due to the above interactions, and its corresponding diagram, are depicted in Fig. \ref{fig:g2-process}. 
For the sake of completeness, we write down the interaction $V_p^{(2)}$ in the second-quantised form
\begin{equation}
V_p^{(2)}=\frac{\tilde{g}_2}{2L}\sum_{Q,\tilde{k},\tilde{q}}\mathrm{sgn}(\tilde{k}-\tilde{q})\mathrm{sgn}(\tilde{k}+\tilde{q})c_{Q+\tilde{k}+\tilde{q}}^{\dagger}c_{Q-\tilde{k}-\tilde{q}}^{\dagger}c_{Q-\tilde{k}+\tilde{q}}c_{Q+\tilde{k}+\tilde{q}},
\end{equation}
where we have relabelled the momenta in the sum as $Q=K/2=(k+k')/2$, $\tilde{k}=(k-k')/2$ and $\tilde{q}=q/2$. 

%%%%%%%%%%%%%Scattering theory%%%%%%%%%%%%%%%%%%%%%%%%%%
\section{Collision-theoretic treatment}
All relevant scattering properties of the system can be extracted from the transition matrix ($T$-matrix) $T(z)$. After separation of centre of mass and relative coordinates, the $T$-matrix for an odd-wave interaction satisfies the following Lippmann-Schwinger equation
\begin{equation}
T(z)=V_p+V_pG_0(z)T(z),
\end{equation}
where $z$ is a complex (relative) energy, and $G_0(z)=[E-H_0]^{-1}$ is the non-interacting Green's function, with $H_0$ the non-interacting relative Hamiltonian. 
%%%%%%%%%%%%%%\subsection{The principal value $T$-matrix}
The scattering states of a Hamiltonian of the form $H=H_0+V$ correspond to an incident wave $\ket{k}$ with non-interacting energy $\epsilon(k)$ and a scattered wave $\ket{\psi_{s,k}^{\pm}}$. This is given by
\begin{equation} 
\ket{\psi_{s,k}^{\pm}}=\lim_{\eta\to 0^{\pm}}G_0(\epsilon(k)+i\eta)T(\epsilon(k)+i\eta)\ket{k}\equiv G_0(\epsilon(k)+i0^{\pm})T(\epsilon(k)+i0^{\pm})\ket{k}.
\end{equation}
Above, the $\pm$ signs correspond to waves propagating forward ($+$) and backwards ($-$) in time, as dictated by time-dependent scattering theory \cite{Joachain}.
The non-interacting Green's function on the energy shell, $G_0(\epsilon(k)+i\eta)$, can be written as
\begin{equation}
\bra{q}G_0(\epsilon(k)+i0^{\pm})\ket{q'}= 2\pi \delta(q-q')\left[\mathcal{P}\left(\frac{1}{\epsilon(k)-\epsilon(q)}\right) \mp i \pi \delta(\epsilon(k)-\epsilon(q))\right],
\end{equation}
where $\mathcal{P}$ denotes Cauchy's principal value. We now define the principal value $T$-matrix $\mathcal{T}$ and the principal value non-interacting Green's function $\mathcal{G}_0$ as
\begin{align}
\mathcal{G}_0(\epsilon(k))&=\frac{1}{2}\left[G_0(\epsilon(k)+i0^+)+G_0(\epsilon(k)-i0^-)\right],\\
\mathcal{T}(\epsilon(k))&=\frac{1}{2}\left[T(\epsilon(k)+i0^+)+T(\epsilon(k)-i0^-)\right].
\end{align}
It is easy to see that the principal value $T$-matrix \footnote{This is also called reaction matrix ($R$-matrix) in the literature.} for real positive energies $E$ satisfies the Lippmann-Schwinger equation
\begin{equation}
\mathcal{T}(E)=V_p+V_p\mathcal{G}_0(E)\mathcal{T}(E).\label{PVT}
\end{equation}
The advantage of using the principal value $T$-matrix is that it avoids the use of imaginary -- and irrelevant except for bound states -- parts, and it completely determines the scattering properties of a system of two identical particles.

\subsection{Interbranch scattering: quadratic dispersion}
We now solve the Lippmann-Schwinger equation, Eq. (\ref{PVT}), for the interaction $V_p^{(2)}$ in Eq. (\ref{Vp2}). The $T$-matrix is readily calculated and reads
\begin{equation}
\mathcal{T}(E;k,k')=\tilde{g}_2\mathrm{sgn}(k)\mathrm{sgn}(k').
\end{equation}
After straightforward but tedious integrations, we find the position-represented fermionic wave functions in the relative coordinate $x$ for relative momentum $k$ and energy $\hbar^2k^2/2\mu$ ($\mu=m/2$ is the reduced mass),
\begin{equation}
\psi_k(x)=\left[1-\frac{m \tilde{g}_2}{\pi\hbar^2|k|}\mathrm{Ci}(|kx|)\right]\sin(kx)-\frac{m \tilde{g}_2}{\pi\hbar^2|k|}\mathrm{Si}(kx)\cos(kx).\label{wf}
\end{equation}
Above, $\mathrm{Ci}$ ($\mathrm{Si}$) is the cosine (sine) integral. The only relevant information in the above wave function is the asymptotic ($x\to \pm \infty$) behaviour. At long distances, the wave function in Eq. (\ref{wf}) behaves as
\begin{equation}
\psi_k(x)\propto \sin(kx)-\frac{m\tilde{g}_2}{2\hbar^2k}\mathrm{sgn}(x)\cos(kx).
\end{equation}
We compare the above asymptotic wave function with the general asymptotic form of a fermionic scattering state in 1D, $\psi_k(x)\propto \mathrm{sgn}(x)\sin(k|x|+\theta_k)$, and we immediately identify the relation between the coupling constant $\tilde{g}_2$ and the scattering phase shift $\theta_k$, as
\begin{equation}
\tan \theta_k = -\frac{m\tilde{g}_2}{2\hbar^2k}.
\end{equation}
We still need to decide upon a renormalisation condition, that is, the collisional energy at which we wish to renormalise the interaction, in which case the phase shift at that specific energy is exact. Naturally, we should renormalise the interbranch scattering phase shift for particles near opposite Fermi points $\pm k_F$, that is, with relative momentum $k_F$. This yields the following value for the bare coupling constant
\begin{equation}
\tilde{g}_2=-\frac{2\hbar^2k_F}{m}\tan\theta_{k_F}=-2\hbar v_F \tan \theta_{k_F},\label{g2quad}
\end{equation}
where we have defined the Fermi velocity $v_F=\hbar k_F/m$.

%%%%%%%%%%%%%%%\subsection{Renormalisation with the linearised dispersion}
\subsection{Interbranch scattering: linearised dispersion}
We now consider the approximate linearised spectrum around the Fermi momenta $\pm k_F^{(0)}$, which is the starting point for building Luttinger liquid theories for fermions. We have included the $^{(0)}$ superscript since we would like to treat, in general, the Fermi points as {\it bare} quantities of the theory. This might sound unusual, but notice that (i) upon {\it any} approximation all coupling constants of the theory may need to be renormalised and in this case it is the dispersion itself that has been expanded (and truncated) as a power series; and (ii) renormalisation of the Fermi points is {\it always} implicitly assumed in Luttinger liquid phenomenology \cite{Giamarchi}, as we shall see below.  

The single-particle energy dispersion is well approximated by
\begin{equation}
\epsilon(k)\approx E_F^{(0)}+\hbar v_F (|k|-k_F^{(0)}),\label{lineardispersion}
\end{equation}
provided that $|k-k_F^{(0)}|/k_F^{(0)}\ll 1$. In Eq. (\ref{lineardispersion}) we have defined the (bare) Fermi energy $E_F^{(0)}=\hbar^2[k_F^{(0)}]^2/2m$. The many-fermion system with the linearised dispersion (\ref{lineardispersion}) corresponds to Tomonaga's model \cite{Tomonaga}, where $k\in (-\infty,\infty)$. In order to connect with the phenomenological model later, we will call right (left) movers particles with $k>0$ ($k<0$). For interbranch collisions we need the relative energy $\epsilon_{\mathrm{rel}}$ for a right and a left mover with relative momentum $k$. This is given by
\begin{equation}
\epsilon_{\mathrm{rel}}(k)=-2E_F^{(0)}+2\hbar v_F |k|.
\end{equation}
For the interaction ($\ref{Vp2}$), the principal value $T$-matrix is readily calculated. Notice that the inverse $T$-matrix is ultraviolet (UV) divergent, and we therefore place a momentum cutoff $\Lambda$. The bare $T$-matrix takes the form $\mathcal{T}(E;k,k')=\tau(E)\mathrm{sgn(k)}\mathrm{sgn}(k')$, with
\begin{equation}
\tau(\epsilon_{\mathrm{rel}}(k))=\frac{1}{\frac{1}{\tilde{g}_2}-\frac{1}{2\pi\hbar v_F}\log \left|\frac{\Lambda-|k|}{|k|}\right|}.\label{invT}
\end{equation}
We now have two options. We can renormalise either the interaction coupling constant $\tilde{g}_2$ while letting the Fermi momentum take its physical value ($k_F^{(0)}=k_F$), or we may renormalise the bare Fermi momentum $k_F^{(0)}$. The renormalisation of the interaction coupling constant is straightforward. Applying the renormalisation condition $\tau(\epsilon_{\mathrm{rel}}(k_F))=g_2^{(R)}$, we obtain
\begin{equation}
\frac{1}{\tilde{g}_2}=\frac{1}{g_2^{(R)}}+\frac{1}{2\pi\hbar v_F} \log \left|\frac{\Lambda - k_F}{k_F}\right|,
\end{equation}
and then we have 
\begin{equation}
\tau(\epsilon_{\mathrm{rel}}(k)) = \frac{1}{\frac{1}{g_2^{(R)}}+\frac{1}{2\pi \hbar v_F}\log\left|\frac{k}{k_F}\right|}.
\end{equation}
The inverse $T$-matrix is now well-defined at almost all energies. It still retains an infrared (IR) divergence, which implies low-energy scattering is suppressed in this approximation. However, this is unimportant, since we will only be concerned with high-energy (of $\mathcal{O}(2E_F)$) processes.

The other possibility, namely the renormalisation of the Fermi points, is quite easy to implement, too. Notice that for $\tilde{g}_2$ to be finite (and equal to $g_2^{(R)}$), all we need is $k_F^{(0)}=\Lambda/2+k_F$, with $k_F$ the renormalised (i.e. physical) Fermi momentum. In this case, the inverse $T$-matrix, Eq. (\ref{invT}), at energy $2E_F^{(0)}$, is regular and equal to $\tilde{g}_2$ when $\Lambda \to \infty$. The $T$-matrix then takes the form
\begin{equation}
\tau(\epsilon_{\mathrm{rel}}(k)) = \frac{1}{\frac{1}{g_2^{(R)}}-\frac{1}{2\pi \hbar v_F} \log\left|\frac{\Lambda}{|k|}-1\right|}.
\end{equation}
Written in the above form, the $T$-matrix appears to be ill-defined. However, if we are interested in relative momenta around the (bare) Fermi points, $k=\Lambda/2 + \tilde{k}$, with $\tilde{k}=\mathcal{O}(k_F)$, we have a constant $T$-matrix $\tau(\epsilon_{\mathrm{rel}}(k))=g_2^{(R)}$ as we take the cutoff to infinity. For momenta far away from the (bare) Fermi points, we get a logarithmically suppressed effective interaction instead. 

All we need to do now is to obtain the renormalisation condition in terms of the phase shift at $k=k_F$. In order for the renormalisation of the Fermi points to make sense, we need to assign the value of the phase shift at the physical value of $k_F$ to scattering at energy $2E_F^{(0)}$. From the asymptotic analysis of the scattering wave function we obtain the renormalisation condition
\begin{equation}
g_2^{(R)} = -2\hbar v_F \tan \theta_{k_F},\label{g2R}
\end{equation}
which is exactly the same renormalisation prescription as for the quadratic dispersion, Eq. (\ref{g2quad}), showing that $\tilde{g}_2$ is unrenormalised when passing from the quadratic to the linear approximation. 

The renormalisation of the Fermi points, while leaving the interaction coupling constant be renormalised without infinities, naturally leads to Luttinger's model. To see this, shift the momentum of right (left) movers as $k\to k+\Lambda/2$ ($k-\Lambda/2$), and rescale the cutoff as $\Lambda/2\to \Lambda$. Then, the momenta of right and left movers both belong to the interval $(-\Lambda,\Lambda)$. The non-interacting Hamiltonian becomes
\begin{equation}
H_0-\mu_0N=\hbar v_F \sum_{k=-\Lambda}^{\Lambda}\left[c_{kR}^{\dagger}kc_{kR}-c_{kL}^{\dagger}kc_{kL}\right],\label{Luttingernonint}
\end{equation}
where $\mu_0=E_F^{(0)}=\hbar^2v_F\Lambda/4+E_F$ and $N=N_R+N_L$ is the total number of particles, and where we have defined $c_{kR}$ ($c_{kL}$) as the annihilation operator for a right (left) mover with momentum $k+\Lambda$ ($k-\Lambda$). The above Hamiltonian, Eq. (\ref{Luttingernonint}), exactly corresponds, after taking the cutoff $\Lambda\to \infty$, to the non-interacting Luttinger Hamiltonian. This shows that, firstly, the renormalisation of the Fermi points is a usual, implicit assumption in most treatments of Luttinger liquid phenomenology and, secondly, that the interaction coupling constant remaining "unrenormalized" \cite{Giamarchi} (that is, renormalised without infinities), is a consequence of the renormalisation of the Fermi points, and taking the limit $\Lambda \to \infty$ {\it before} allowing the coupling constant to change at all.
%%%%%%%%%%%%%%%g4-processes%%%%%%%%%%%%%%%%%%%%%
\section{Intrabranch processes} 
The intrabranch coupling constant, typically denoted by $g_4$ \cite{Giamarchi}, is very simple to obtain as a function of the interbranch coupling constant, since it must obey a natural physical constrain, corresponding to the conservation of charge (particle number) \cite{BruusFlensberg}. In our notation, it corresponds to setting $g_4 = 2g_2^{(R)}$. However, the understanding of the microscopic origin of $g_4$ is far from satisfactory, and we study this in detail in this section.

\begin{figure}[t]
\includegraphics[width=1\textwidth]{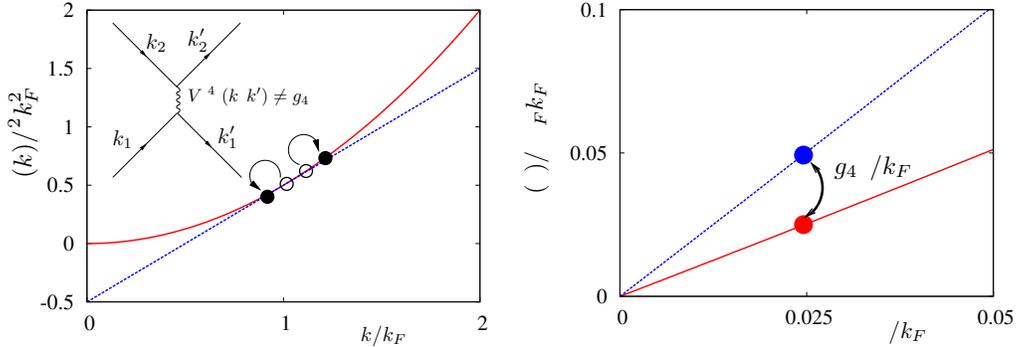}
\caption{Left: depiction and interaction diagram for intrabranch collisions. The red solid line corresponds to the Galilean (parabolic) dispersion, and the blue dashed line corresponds to the linearised approximation around the Fermi points. Right: depiction of the effect of the effective intrabranch interaction ($g_4$), to raise the excitation energies of quasi-particles (see text).}
\label{fig-g4}
\end{figure}

It is customary to consider intrabranch interactions as two-particle scattering processes \cite{Giamarchi, Eggert} (see Fig. \ref{fig-g4}), and then use a phenomenological coupling constant, namely $g_4$, that simply adds to the Fermi velocity and renormalises the bosonic excitation energies in the Luttinger Liquid \cite{Giamarchi}. There are, however, several problems associated with this collisional interpretation of intrabranch processes. Firstly, before linearising the single-particle dispersion, one can define two-fermion scattering states in, say, the right-moving branch (we will consider right-moving fermions throughout this section unless otherwise stated). Since the relevant scattering processes occur around the Fermi point, the energy at which one should renormalise the interactions should be twice the Fermi energy $2E_F$. However, the fermionic $T$-matrix (ergo the phase shift) tends to zero quadratically with relative momentum, see Eq. (\ref{V4}). This would mean that the renormalised Fermi velocity would not be changed to zero-th order in the momentum of excitations, contrary to well-known results and phenomenology \cite{Giamarchi}. The second, even more severe problem, is that even if we wanted to renormalise the scattering properties in the linearised approximation to match the exact scattering properties (obtained using the full quadratic dispersion) around the Fermi point, we would simply not be able to do so. The fact, seemingly unnoticed so far, is that two particles (fermions or bosons) in the same branch of a linearised dispersion do not scatter. We begin by considering this issue in the following subsection.

\subsection{Absence of collisions}
Consider the Lippmann-Schwinger equation off the energy shell $E\ne 0$ in the relative coordinate,
\begin{equation}
\bra{k'}\mathcal{T}(E)\ket{k}=V(k,k')+\frac{1}{2\pi}\mathcal{P}\int_0^{\infty} dq\frac{V(k,q)}{E}\bra{q}\mathcal{T}(E)\ket{k'},
\end{equation}
where $\ket{k}$ is regarded here as a fermionic (antisymmetric) state, that is $2^{-1/2}(\ket{k}-\ket{-k})\to \ket{k}$. As we go on-shell, $E\to 0$, we see that the principal value $T$-matrix must vanish linearly with $E$. Therefore, we define $\mathcal{T}(E)=E t(0)$ for $E\to 0$. As we take the limit $E\to 0$, we obtain the following integral equation
\begin{equation}
V(k,k')=-\frac{1}{2\pi}\int_{0}^{\infty}dqV(k,q)\bra{q}t(0)\ket{k'}.
\end{equation}
The above equation can be written as an operator identity, which reads
\begin{equation}
-V=Vt(0).\label{opeq}
\end{equation}
The simplest way to solve Eq. (\ref{opeq}) is by diagonalising $V$. We define a unitary operator $U$ \footnote{Since $V$ can be quite degenerate, the operator $U$ is not necessarily unique.} that diagonalises $V$, together with its associated orthogonal set of eigenfunctions $\ket{\alpha_k}$ and their corresponding eigenvalues $V(\alpha_k)$, such that
\begin{equation}
\ket{\alpha_k}=U\ket{k}.
\end{equation}
Upon transformation of $V$ and $t(0)$ with $U$, Eq. (\ref{opeq}) reduces to
\begin{equation}
-2\pi V(\alpha_q) \delta(q-q') = V(\alpha_{q'}) \bra{\alpha_{q'}}t(0)\ket{\alpha_q}.\label{equalpha}
\end{equation}
For a strict finite-range potential $V$ (which we can always consider as a good approximation), most of its eigenvalues will be degenerate and equal to zero. We must therefore deal with Eq. (\ref{equalpha}) with care. Let us define the set of values of $q$ for which $V(\alpha_q)=0$, that we define as $I_0$, that is
\begin{equation}
I_0=\{q\ge 0:V(\alpha_q)=0\},
\end{equation}
while we denote its complementary set by $I_V\equiv 1-I_0$. If $q\in I_0$, then for each $q'$ we must have either $q'\in I_0$ or $\bra{\alpha_{q'}}t(0)\ket{\alpha_q}=0$. If $q' \in I_0$ as well, then from Eq. (\ref{equalpha}) we see that $\bra{\alpha_{q'}}t(0)\ket{\alpha_q}$ is arbitrary, which is a natural consequence of degeneracy. In this case, we shall fix its value to a constant $\omega$, i.e.
\begin{equation}
\bra{\alpha_{q'}}t(0)\ket{\alpha_q} = \omega, \hspace{0.1cm} q,q' \in I_0.
\end{equation}
For $q'\in I_V$, we are obviously forced to set $\bra{\alpha_{q'}}t(0)\ket{\alpha_q}=0$. We now consider the case $q\in I_V$. From Eq. (\ref{equalpha}) we have that if $q\ne q'$, then $\bra{\alpha_{q'}}t(0)\ket{\alpha_q}$ is arbitrary and we set it to the constant $\omega$ as above if $q' \in I_0$. The case $q=q'$ yields $\bra{\alpha_q}t(0)\ket{\alpha_q}=-2\pi$. 

In the discussion above we have completely solved the Lippmann-Schwinger equation. We still need to calculate the scattering states. These have the form
\begin{equation}
\ket{\psi}=\ket{k}+\lim_{E\to 0}\mathcal{G}^{(0)}(E)\mathcal{T}(E)\ket{k}.
\end{equation}
Inserting a resolution of the identity, we find the total scattering state as
\begin{equation}
\ket{\psi}=\ket{k}+\int_0^{\infty}\frac{dk'}{2\pi}\ket{k'}\left[\omega \int_{I_0}\frac{dq'}{2\pi}\langle k'|\alpha_{q'}\rangle\right.\left.\int_{0}^{\infty}\frac{dq'}{2\pi}\langle \alpha_q | k\rangle - \int_{I_V} \frac{dq}{2\pi} \langle k' | \alpha_q\rangle \langle \alpha_q | k\rangle\right].\label{totscat}
\end{equation}
It is very illustrative to choose $\omega=0$ above. Using the resolution of the identity $1$ in Eq. (\ref{totscat}), and identifying
\begin{equation}
1-P_{I_0} = \int_{I_V} \frac{dq}{2\pi} \ket{\alpha_q}\bra{\alpha_q},
\end{equation}
where $P_{I_0}$ is the projector onto $I_0$, we find that
\begin{equation}
\ket{\psi}=P_{I_0}\ket{k}.\label{projector}
\end{equation}
The above result has a clear interpretation: the scattering state of the system associated with the incident wave $\ket{k}$ is nothing but the projection of the incident wave onto the zero potential eigenstates. An immediate consequence of Eq. (\ref{projector}) is that our system (or any other flat-banded system) only supports scattering states if the potential has at least one vanishing eigenvalue. 

The above analysis proves our claim that intrabranch interactions can lead to either (i) energy shifts if the potential has no zero eigenvalues or (ii) a superposition of non-interacting waves that avoid potential overlap if the interaction has vanishing eigenvalues \footnote{These correspond to the "strange solutions" of one-dimensional field theories found by Mattis and Sutherland \cite{MattisSutherland}.}. 

\subsection{Fermion-fermion intrabranch interactions}\label{Section52}
It is clear from the discussion in the previous subsection that the $g_4$-processes do not correspond to particle-particle intrabranch scattering. Luttinger liquid theory, on the other hand, predicts that particle-hole excitation energies are shifted due to intrabranch interactions. We will see in the following that studying the energy shifts of particle-hole excitations in the chiral (right- and left-moving) branches separately allows for some sort of renormalisation of the $g_4$ coupling constant in the weak {\it inter}-branch interaction regime. We will show, however, that microscopic fermion-fermion intrabranch interactions {\it do not} correspond to the intrabranch interactions appearing in Luttinger liquid phenomenology. These will be considered in the next subsection.

We begin by defining the chiral right-moving problem, with single-branch Hamiltonian $H_{\mathrm{R}}$ given by
\begin{equation}
H_{\mathrm{R}}=\sum_{k>0} \epsilon(k)c_k^{\dagger}c_k + V_{\mathrm{RR}},\label{HRR}
\end{equation}
where 
\begin{equation}
V_{\mathrm{RR}} = \frac{1}{2L}\sum_{k,k'>0,q}V(q) c_{k+q}^{\dagger}c_{k'-q}^{\dagger}c_{k'}c_k \theta(k+q)\theta(k'-q).\label{VRR1}
\end{equation} 
Hamiltonian (\ref{HRR}) conserves total momentum. Since the Fermi sea $\ket{N_R}$ for $N_R$ right-moving fermions, given by
\begin{equation}
\ket{N_R}=\prod_{0<k\le k_F}c_k^{\dagger}\ket{0},
\end{equation}
is the unique state with lowest total momentum, it is an eigenstate of the interacting Hamiltonian $H_{\mathrm{R}}$. For quite general repulsive interactions we can also assume it is the ground state of the chiral system. In order to obtain particle-hole excitation energies, it is convenient to rewrite the interaction (\ref{VRR1}) in the following way (throwing away constant energy shifts)
\begin{equation}
V_{\mathrm{RR}}=\frac{1}{L}\sum_{q>0}V(q) \rho_q^{+}\rho_q -\frac{1}{L} \sum_{kq>0} V(q)c_{k+q}^{\dagger}c_{k+q}\label{VRR2}
\end{equation} 
Above, we have defined the displacement operators $\rho_{qR}$ and $\rho_{qR}^{+}$ as
\begin{align}
\rho_{qR}&=\sum_{k>q}c_{k-q}^{\dagger}c_{k}\\
\rho_{qR}^{+}&=\sum_{k>0}c_{k+q}^{\dagger}c_k = \rho_{-q}.
\end{align}
Since the first term on the r.h.s. of Eq (\ref{VRR2}) is bosonizable, we shall first deal with the second term, which we denote by $\mathcal{V}$. In particular, we are interested in the difference in energy between particle-hole excitations and the Fermi sea due to this interaction term. We will denote particle-hole states by $\ket{\tilde{k}+Q;\tilde{k}}$, with
\begin{equation}
\ket{\tilde{k}+Q;\tilde{k}}=c_{\tilde{k}+Q}^{\dagger}c_{\tilde{k}}\ket{N_R},
\end{equation}
where obviously $\tilde{k}\le k_F$ and $Q>0$. The action of $\mathcal{V}$ on a particle-hole state, throwing away a trivial constant term so that it gives the excitation energies, is given by
\begin{equation}
\mathcal{V}\ket{\tilde{k}+Q;\tilde{k}}=\frac{1}{L}\left[\sum_{k=0}^{\tilde{k}}V(\tilde{k}-k)-\sum_{k=0}^{\tilde{k}+Q}V(\tilde{k}+Q-k)\right]\ket{\tilde{k}+Q;\tilde{k}},
\end{equation}
which means particle-hole states are eigenvectors of $\mathcal{V}$. Since we are interested in Luttinger liquid phenomenology, we will consider only excitation energies of $\mathcal{O}(Q)$. To this order, we can obtain these analytically by expanding $V(\tilde{k}+Q-k)$ as
\begin{equation}
V(\tilde{k}+Q-k)=V(\tilde{k}-k)+\frac{dV}{d\kappa}{\bigg |}_{\kappa=\tilde{k}-k}Q+\mathcal{O}(Q^2).
\end{equation}
We obtain
\begin{equation}
\mathcal{V}\ket{\tilde{k}+Q;\tilde{k}}=\left\{\frac{1}{2\pi}\left[V(Q)-V(\tilde{k})\right]Q\right.\left.-\frac{1}{2\pi}\int_{\tilde{k}}^{\tilde{k}+Q}dk V(\tilde{k}-k)\right\}\ket{\tilde{k}+Q;\tilde{k}}+\mathcal{O}(Q^2).
\end{equation}
In order to obtain the leading order in $Q$ above, we set $V(Q)=V(0)+\mathcal{O}(Q^2)$ and finally arrive at
\begin{equation}
\mathcal{V}\ket{\tilde{k}+Q;\tilde{k}}=\left[-\frac{1}{2\pi} V(\tilde{k})Q + \mathcal{O}(Q^2)\right]\ket{\tilde{k}+Q;\tilde{k}}.
\end{equation}
Above, we see that the excitation energy to linear order in $Q$ due to the part $\mathcal{V}$ of the intrabranch interaction is given by $-V(\tilde{k})Q/2\pi$.

We now consider the first term on the r.h.s. of Eq. (\ref{VRR1}), which we will denote by $\mathcal{V}_B$, i.e. the part of the interaction that is straightforwardly bosonizable. For sufficiently small $q$ ($q\ll k_F$), the usual commutation relations for the displacement operators \cite{Giamarchi} are valid,
\begin{equation}
[\rho_{qR},\rho_{q'R}^+]=\frac{Lq}{2\pi}\delta_{q,q'}.
\end{equation}
We define bosonic operators $a_{qR}=(2\pi/Lq)^{1/2}\rho_{qR}$ (and equivalently for $a_{qR}^{\dagger}$), and the term $\mathcal{V}_B$ is rewritten as
\begin{equation}
\mathcal{V}_B=\frac{1}{2\pi}\sum_{q>0}qV(q) a_{qR}^{\dagger}a_{qR}.
\end{equation}
The eigenstates of $\mathcal{V}_B$ including only single particle-hole pairs are given by
\begin{equation}
a_{qR}^{\dagger}\ket{N_R}=\left(\frac{2\pi}{Lq}\right)^{1/2}\sum_{k_F-q<k\le k_F}\ket{k+q;k},
\end{equation}
and their respective eigenvalues are given by $qV(0) + \mathcal{O}(q^2)$.

In order to establish the excitation energies solely in terms of a quadratic bosonic Hamiltonian, we need to make two approximations at this point. One of them is the linearisation of the single-particle dispersion, so that the excitation kinetic energy of a particle-hole state $\ket{\tilde{k}+Q;\tilde{k}}$, which is given by $\hbar^2\tilde{k}Q/m +\mathcal{O}(Q^2)$, is approximated by $\hbar v_FQ/m$ for all $\tilde{k}$. The second one corresponds to approximating the energy shift due to $\mathcal{V}$ as
\begin{equation}
-\frac{1}{2\pi}V(\tilde{k})Q\approx -\frac{1}{2\pi}V(k_F)Q,
\end{equation}
for all $\tilde{k}$, a reasonable approximation for small $Q$. In that approximation, $a_{qR}^{\dagger}\ket{N_R}$ are also eigenstates of $\mathcal{V}$, and we can finally write down the intrabranch interaction, approximately, as
\begin{equation}
V_{\mathrm{RR}}=\frac{1}{2\pi}\sum_{q>0} q \left[V(0)-V(k_F)\right]a_{qR}^{\dagger}a_{qR}.
\end{equation}
The above intrabranch interaction would give a (weak-coupling) value of $g_4=V(0)-V(k_F)$. This is incorrect, since in the weak-coupling limit the result should be $g_4=V(0)-V(2k_F)$. What has gone wrong here is that the intrabranch interaction in the {\it effective Luttinger's model} does not correspond to an intrabranch process in the original fermionic system. We will see how this is the case in the following subsection.

\subsection{Luttinger liquid intrabranch interaction from the original interbranch process}
We consider now the weak-coupling limit of the intrabranch interaction. The relevant state to be considered is a particle-hole excitation from the static Fermi sea in the original fermionic system, 
\begin{equation}
\ket{\tilde{k}+Q}\equiv c_{\tilde{k}+Q}^{\dagger}c_{\tilde{k}}\ket{F},
\end{equation}
where $0<\tilde{k}\le k_F$ and $0<Q\ll k_F$, while
\begin{equation}
\ket{F}=\prod_{k=-k_F}^{k_F}c_{k}^{\dagger} \ket{0}.
\end{equation}
We denote the relevant part of the interaction by $V'$, which is given by
\begin{equation}
V'=\frac{1}{L}\sum_k\left[V(0)-V(k-\tilde{k}-Q)\right]c_k^{\dagger}c_k c_{\tilde{k}+Q}^{\dagger}c_{\tilde{k}}.
\end{equation}
All other interaction processes give contributions of $\mathcal{O}(Q^2)$ or higher. We obtain 
\begin{equation}
V'\ket{\tilde{k}+Q}=\frac{1}{2\pi}\int_{-k_F}^{k_F} dk \left[V(0)-V(k-\tilde{k}-Q)\right] \ket{\tilde{k}+Q,\tilde{k}}.
\end{equation}
Subtracting the energy shift corresponding to the Fermi sea, i.e. that obtained by setting $Q=0$ above, and after using $\tilde{k}\approx k_F$, we obtain the energy shift $g_4^{(1)}Q/2\pi$ to $\mathcal{O}(Q)$, with
\begin{equation}
g_4^{(1)} = V(0)-V(2k_F),
\end{equation}
which is the correct answer we were looking for. 

\section{Final Hamiltonian and Luttinger parameters}
With the notations used in the previous sections, we can write down the effective Luttinger liquid Hamiltonian of the system in bosonized form, i.e. in terms of the boson operators for right (left) movers $a_{qR}$ ($a_{qL}$). This is given by
\begin{equation}
H=\hbar v\sum_{q>0}q\left[a_{qR}^{\dagger}a_{qR}+a_{qL}^{\dagger}a_{qL}\right]+\frac{g_2}{2\pi}\sum_{q>0}q\left[a_{qR}^{\dagger}a_{qL}^{\dagger}+a_{qL}a_{qR}\right],\label{Luttingerbosonic}
\end{equation}
where $v=v_F+g_4/2\pi\hbar$, and where $g_2=2g_2^{(R)}$, see Eq. (\ref{g2R}). The last statement might not be so easy to grasp, so we prove it as follows, using the notation of ref. \cite{BruusFlensberg}. The original interbranch interaction is given by
\begin{align}
V^{(2)}&=\frac{1}{2L} \sum_{k>0,k'<0,q\ne 0} \left[V(q)-V(k'-k-q)\right] \nonumber \\
&\times c_{kR}^{\dagger}c_{k'+q,R}c_{k',L}^{\dagger} c_{k'-q,L} +(\mathrm{R}\leftrightarrow \mathrm{L}).\label{RLthing}
\end{align}
To see how the p-wave interaction arises due to the exchange above, we define $Q=(k-k')/2+q$ and $Q'=(k-k')/2$, in which case $q=Q-Q'$ and $k'-k-q=-(Q+Q')$. Therefore, we have $V(q)-V(k'-k-q)=2V_p(Q,Q')$. Replacing now the p-wave potential by the "constant" interaction, we obtain
\begin{equation}
V_p(Q,Q') = \tilde{g}_2 \mathrm{sgn} \left(\frac{k-k'}{2}+q\right)\mathrm{sgn}\left(\frac{k-k'}{2}\right).
\end{equation}
Then for the first term in Eq. (\ref{RLthing}) we have $(k-k')/2\approx k_F>0$ and, for small $q$ (forward scattering), we have $k_F+q>0$, and therefore $V_p(Q,Q')\approx \tilde{g}_2$, which leads to the $g_2$-interaction term in Eq. (\ref{Luttingerbosonic}).

Hamiltonian (\ref{Luttingerbosonic}) corresponds to the notation in ref. \cite{Giamarchi}, and the Luttinger parameter $\mathcal{K}$ and the speed of excitations $u$ are given by
\begin{align}
\mathcal{K}&=\left[\frac{1+\frac{g_4}{2\pi \hbar v_F}-\frac{g_2}{2\pi\hbar v_F}}{1+\frac{g_4}{2\pi \hbar v_F}+\frac{g_2}{2\pi\hbar v_F}}\right]^{1/2},\label{LuttingerK}\\
u&=v_F\left[\left(1+\frac{g_4}{2\pi \hbar v_F}\right)^2-\left(\frac{g_2}{2\pi\hbar v_F}\right)^2\right]^{1/2}.\label{Luttingeru}
\end{align}
In the weak-coupling limit, we can expand the Luttinger parameter and the speed of excitations as follows
\begin{align}
\mathcal{K} &= 1 -\frac{g_2}{2\hbar \pi v_F}+\mathcal{O}(g_2^2),\\
u &= v_F\left(1+\frac{g_4}{2\hbar \pi v_F}\right) + \mathcal{O}(g_2^2).
\end{align}
Our simple renormalisation prescription gives the first order results above exactly. We have to notice that the first order results in $g_2$ (or equivalently $g_4$), obtained from our renormalisation prescription are not perturbative, since they represent the exact two-body phase-shifts of the system and not their first Born approximation $\propto V(0)-V(2k_F)$.

\section{Lieb-Liniger and Cheon-Shigehara models}
We apply now the results of the previous sections to a particular example of an exactly solvable model, namely the (bosonic) Lieb-Liniger gas \cite{LiebLiniger}, or its fermionic dual, the Cheon-Shigehara model \cite{CheonShigehara}. 

For the sake of clarity, and to avoid confusion with other notations in the literature, we write down the Lieb-Liniger Hamiltonian $H_{\mathrm{Lieb}}$ in first quantised form
\begin{equation}
H_{\mathrm{Lieb}}=\sum_i \frac{p_i^2}{2m}+g\sum_{i<j} \delta(x_i-x_j).
\end{equation}
The dual fermion-fermion interaction in Cheon-Shigehara's model, however, has a formidable form in the position representation. Its momentum representation, on the other hand, is very simple, and does not appear to be widely known. This is given in ref. \cite{ValienteZinner}, and has the form
\begin{equation}
V_{\mathrm{CS}}(k,k') = g_F kk',\label{VCS}
\end{equation}
which corresponds to a p-wave interaction (see Eq. (\ref{pwave})). From the interaction in Eq. (\ref{VCS}), we obtain the following phase shift \cite{ValienteZinner}
\begin{equation}
\tan \theta_{\mathrm{CS}} = -\frac{mg_F}{2\hbar^2}k.
\end{equation}
The dual interaction in Lieb-Liniger's model is obtained by equating the $\theta_{\mathrm{CS}}$ to the Lieb-Liniger phase shift \cite{LiebLiniger}, obtaining
\begin{equation}
g_F = -\frac{4}{g}\left(\frac{\hbar^2}{m}\right)^2.
\end{equation}
The above discussion immediately yields the value of $g_2$ (and $g_4$) in terms of the Lieb-Liniger coupling constant $g$,
\begin{equation}
g_2=-\frac{8}{g}(\hbar v_F)^2.\label{g2LL}
\end{equation}
Defining now the dimensionless Lieb-Liniger parameter $\gamma$,
\begin{equation}
\gamma = \frac{\pi mg}{k_F \hbar^2} =\frac{\pi g}{\hbar v_F},
\end{equation}
we obtain the desired first order result in the Tonks-Girardeau limit
\begin{align} 
\mathcal{K} &= 1+\frac{4}{\gamma} +\mathcal{O}(\gamma^{-2})\\
u &= 1 - \frac{4}{\gamma} + \mathcal{O} (\gamma^{-2}) ,
\end{align}
in agreement with the known result (see, e.g. \cite{Imambekov}).

It is worth noting that in this model {\it only}, the first Born approximation to $g_2$ is identical to the result using the phase shift. This is due to the fact that Lieb-Liniger's model is a zero-range $S$-matrix theory, i.e. the interaction has zero range and the model is solvable via the Bethe ansatz. In collision-theoretic language, the interaction potential in this model has already been renormalised to give the correct phase shift. To see this, we note that the following interaction has the same action on {\it fermions} as the Cheon-Shigehara interaction (\ref{VCS}),
\begin{equation}
V(q)=-\frac{g_F}{2}q^2,
\end{equation}
which implies $V(0)-V(2k_F)=-(8/g)(\hbar v_F)^2$, which is identical to the result obtained using the phase-shift, Eq. (\ref{g2LL}).

\begin{figure}[t]
\includegraphics[width=1\textwidth]{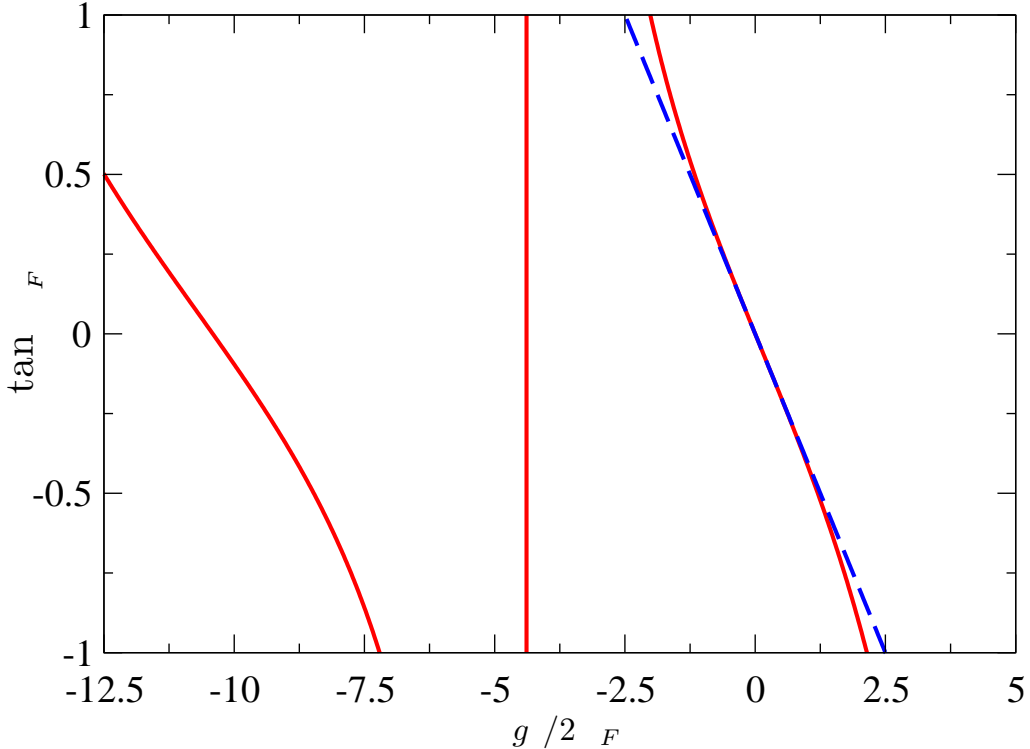}
\caption{The tangent of the phase shift (solid red line) at relative momentum $k=k_F$ for two fermions interacting via an exponential potential, Eq. (\ref{exppot}), compared to its first Born approximation (dashed blue line), for $\lambda/k_F=1$. The vertical red line denotes the position of a shape resonance.}
\label{fig:phaseshift}
\end{figure}

We illustrate now the fact that the first Born approximation to $g_2$ being identical to the non-perturbative result only occurs in Lieb-Liniger's model, where the interaction has been renormalised already for that purpose. We consider a full-blown two-particle interaction of the following form in the position representation
\begin{equation}
W(x)=g_0e^{-\lambda |x|},\label{exppot}
\end{equation}
where $x$ is the relative coordinate, $g_0$ is the strength of the interaction, and $\lambda >0$ controls its range. For two fermions, the phase shift is readily calculated from the relation
\begin{equation}
\tan \theta_k = -\frac{m}{2\hbar^2 k} \int_{-\infty}^{\infty} dy W(y) \sin(ky) \psi_k(y),
\end{equation}
where the fermionic scattering state $\psi_k(x)$ satisfies the Lippmann-Schwinger equation
\begin{equation}
\psi_k(x)=\sin(kx) + \frac{m}{2\hbar^2 k} \int_{-\infty}^{\infty} dy \sin(k|x-y|) W(y) \psi_k(y).
\end{equation}
The results for the phase shift (which give the value of $g_2$) are shown in Fig. \ref{fig:phaseshift} , where we have set $\lambda/k_F=1$. There, we observe that the weak-coupling result is only valid in a small window of weak interaction strength $g_0/E_F$. However, the gas phase of the system can be weakly-coupled for strong attractive interactions, to the left of the shape resonance in Fig. \ref{fig:phaseshift}. In fact, the strongly-attractive side of Lieb-Liniger and Cheon-Shigehara models (the super-Tonks regime), corresponds to this region in the exponential potential's strength, while the weakly attractive regime in Cheon-Shigehara model corresponds to the weakly attractive regime of the exponential potential, and so does the strongly repulsive (Tonks regime) Lieb-Liniger model. We also note that there is no such thing as {\it repulsive} Cheon-Shigehara model. When $g_F>0$ ($g_2>0$) we still have a strongly-attractive interaction due to renormalisation \cite{ValienteZinner}. Cheon-Shigehara model is everywhere attractive: strongly attractive (there are bound states) for $g_F>0$ ($g_2>0$) and weakly attractive (there are no bound states) for $g_F<0$ ($g_2<0$).

\section{Conclusions}
In this article we have studied how the phenomenological constants of Luttinger liquid theories relate to the microscopic interactions of spin-polarised one-dimensional Fermi gases. We have shown that interbranch interactions rule both phenomenological inter- and intrabranch interactions in Luttinger's model. This has also been shown to arise in a completely natural way from the renormalisation of the Fermi points in Tomonaga's model, upon an appropriate redefinition of the wave numbers for right- and left-movers. As an important side result, our renormalisation programme explains, in very simple terms, the well-known fact of unrenormalisabilty of the interbranch coupling constant. 

Our theoretical framework can be generalised to Luttinger liquids with spin, chiral Luttinger liquids -- where the results of section \ref{Section52} are directly applicable -- and helical Luttinger liquids \cite{Helical}. Many of the concepts studied here can be relevant for the study of non-linear Luttinger liquids \cite{Imambekov}, i.e. when the non-linearity in the kinetic energy dispersion becomes important. 

\section*{Acknowledgements}
M.V. and P.\"O. acknowledge support from EPSRC grant No. EP/J001392/1, L.G.C. acknowledges support from the EPSRC CM-DTC.

\bibliographystyle{unsrt}

\end{document}